A **M**ultidatabase **ExTR**action Pip**E**line (METRE**)** for Facile Cross Validation in Critical Care Research

Wei Liao[1], Joel Voldman[1]

[1]Department of Electrical Engineering and Computer Science, Massachusetts Institute of Technology, Cambridge, USA

**Corresponding author:** Joel Voldman: Department of Electrical Engineering and Computer Science, Massachusetts Institute of Technology, 77 Massachusetts Avenue, Room 36-824, Cambridge, MA 02139, USA. E-mail: voldman@mit.edu. Tel: +1 617 253 2094.

## ABSTRACT

Transforming raw EHR data into machine learning model-ready inputs requires considerable effort. One widely used EHR database is Medical Information Mart for Intensive Care (MIMIC). Prior work on MIMIC-III cannot query the updated and improved MIMIC-IV version. Besides, the need to use multicenter datasets further highlights the challenge of EHR data extraction. Therefore, we developed an extraction pipeline that works on both MIMIC-IV and eICU Collaborative Research Database and allows for model cross validation using these 2 databases. Under the default choices, the pipeline extracted 38766 and 126448 ICU records for MIMIC-IV and eICU, respectively. Using the extracted time-dependent variables, we compared the Area Under the Curve (AUC) performance with prior works on clinically relevant tasks such as in-hospital mortality prediction. METRE achieved comparable performance with AUC 0.723-0.888 across all tasks. Additionally, when we evaluated the model directly on MIMIC-IV data using a model trained on eICU, we observed that the AUC change can be as small as +0.019 or -0.015. Our open-source pipeline transforms MIMIC-IV and eICU into structured data frames and allows researchers to perform model training and testing using data collected from different institutions, which is of critical importance for model deployment under clinical contexts.

## 1. INTRODUCTION

Electronic health record (EHR) data contains rich patient-specific information and holds great promise in developing clinical decision support systems, understanding patient trajectories and ultimately improving the quality of care. To date, in combination of machine learning (ML) methods, EHR data has been leveraged to build prediction systems for diseases such as sepsis [1–4], acute kidney failure [5,6], rheumatoid arthritis [7], etc., develop interpretable algorithms [8,9], or explore the state space to learn what treatments to avoid in data-constrained settings [10]. EHR data is usually contained within a relational database, and most machine learning algorithms cannot directly operate on the raw data. Transforming raw EHR data into ML model-ready inputs requires substantial effort, which includes selecting what variables to be extracted, how to organize irregularly recorded time-dependent variables as well as how to handle missing and outlier data points. McDermott et al. [11] found that only 21% of the papers in the machine learning for health (MLH) field released their code publicly, and that researchers usually extract task-specific data, which impedes comparison across different studies.

Recently, there has also been an increasing concern on the potential sources of harm throughout the machine learning life cycle [12,13]. For instance, representation bias can occur when samples used during model development underrepresent some part of the population, and the model thus subsequently fails to generalize for a subset of the use population. In addition, machine learning models can learn spurious correlations between the data and the target output. For example, the model can potentially link a specific variable recording pattern in the EHR training data instead of relevant clinical physiology with the outcome [14], which could lead to failure when the algorithm is deployed to another clinical site. Therefore, in the model development phase, it is important for researchers to perform model training or validation using EHR data from different sources. As McDermott et al. point out [11], whereas ~80% of computer vision studies and ~58% of natural language processing studies used multiple datasets to establish their results, only ~23% of ML for health (MLH) papers did this. The need to use multicenter datasets further highlights the challenge of EHR data preprocessing, because EHR data is usually archived differently across institutions, which requires design of different query strategies.

The exciting research in MLH is made feasible by the availability of large collections of EHR datasets. Medical Information Mart for Intensive Care (MIMIC) is a pioneer in ensuring safe, appropriate release of EHR data. MIMIC-III is a large, freely-available database comprising over 40000 deidentified records of patients who stayed in the critical care unit at Boston's Beth Israel Deaconess Medical Center between 2001 and 2012 [15]. MIMIC-Extract [16] is a popular open source pipeline for transforming the raw critical care MIMIC-III database into data structures that are directly usable in common time-series prediction pipelines. It incorporates detailed unit conversion, outlier detection and missingness thresholding and a semantically similar feature aggregation pipeline to facilitate the reproducibility of research results that use MIMIC-III. Carrying on with the success of MIMIC-III, MIMIC-IV was introduced in 2020 with a few major changes to further facilitate usability. MIMIC-IV states the source database of each table (`Chartevents, Labevents` etc.) while MIMIC-III organizes the data as a whole. MIMIC-IV also contains more contemporary data recorded using the MetaVision system (instead of the CareVue system) from 2008 - 2019, which makes prior query code built on CareVue redundant; MIMIC-IV's new structure and its accompanying level of detail require new query strategies. COP-E-CAT [17] is a preprocessing framework developed for MIMIC-IV that allows users to specify the time window to aggregate the raw features. Despite that utility and impact of both MIMIC-Extract and COP-E-CAT, they are constrained to operating on MIMIC-III or MIMIC-IV data, and thus do not provide developers with an independent dataset to improve model generalization and reduce bias. FIDDLE [18] is a data extraction pipeline for both MIMIC-III and eICU. eICU Collaborative Research Database [19] is a multicenter database comprising deidentified health data associated with over 200000 admissions to ICUs across the United States between 2014-2015. FIDDLE incorporates important advances, such as circumventing the process of selecting variables and making use of the individual EHR variable distributions. However, for a given model developed using FIDDLE-MIMIC-III, it's still challenging to perform direct model evaluation using FIDDLE-eICU. In contrast to previous works, METRE advances the field with 2 primary contributions:

1. METRE works with the most recent MIMIC-IV database. We extract a diverse set of time-dependent variables: 92 labs and vitals, 16 intervention variables as well as 35 time-invariant variables. Our intervention table expands upon those included in MIMIC-Extract (ventilation, vasopressors, fluid bolus therapies) and further includes continuous renal replacement therapy (CRRT) and 3 types of transfusion procedures as well as antibiotics administration, which could be relevant in a variety of clinical tasks. Our default MIMIC-IV cohort is 38766 patient stays who were admitted to the ICU. We also incorporate flexible user-specified inputs into the pipeline. Users can specify the age range, data missingness threshold, ICU stay length, and specific condition keywords to get their unique cohort.

2. We also developed an eICU database extraction pipeline with each extracted feature mapped onto those extracted from MIMIC-IV. The eICU pipeline was developed using the same outlier removal and missing data imputation strategy as MIMIC-IV. To demonstrate the usability of our pipeline, we performed the following tasks using MIMIC-IV and eICU data separately: 1) hospital mortality prediction, 2) acute respiratory failure (ARF) prediction using 4h or 12h ICU data, 3) shock prediction using 4h or 12h data. Afterwards, for models developed using MIMIC-IV, we performed model testing directly on eICU (or vice versa) without any transfer learning.

## 2. METHODS

### 2.1. Pipeline overview

Figure 1 summarizes the extraction steps for both MIMIC-IV and eICU. The extraction pipeline starts with defining the cohort, where users can specify age range and ICU length of stay (LOS) range on the records to be extracted. Besides, we also provide a few arguments to extract condition-specific cohorts. For instance, such arguments include `sepsis_3`, under which the pipeline will only extract patients that meet sepsis-3 criteria [20] during their ICU stay. Such arguments include `sepsis_3`, `ARF`, `shock`, `COPD`, and `CHF`. The definition for each condition is in SI Section 8. Based on the identification info of the cohort (`subject_id`, `hadm_id`, `stay_id` in MIMIC-IV and `patientunitstayid` in eICU), the pipeline proceeds to extract 3 tables, namely Static, Vital and Intervention.

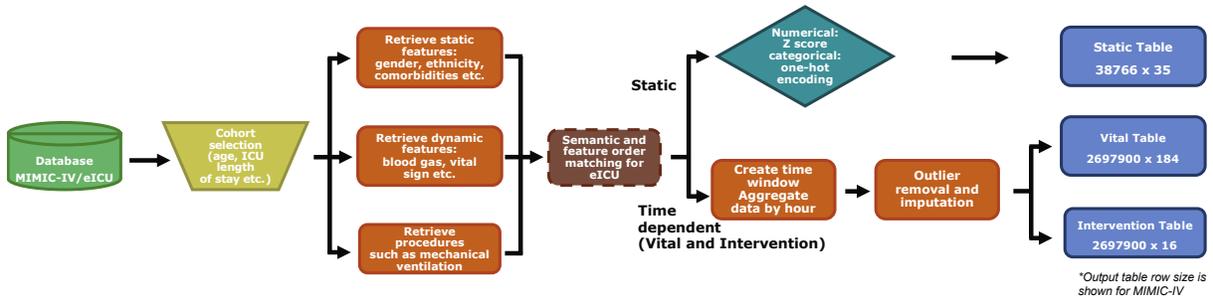

Figure 1. METRE schematic. (Dashed box is specific for eICU. Output dataframe shape is from MIMIC-IV.)

The Static table contains information such as patient age, gender, ethnicity, and comorbidities. Notably, we also extract the 3-year time window for each patient admission in the MIMIC-IV database, since the evolution of care practices over time and the resultant concept drift can significantly change clinical data [21], which can limit model deployability. The release of the approximate admission window is a new feature of MIMIC-IV (versus MIMIC-III). For eICU, all the patients were discharged between 2014 and 2015.

The Vital table contains information such as blood gas and vital sign measurements. These measurements can be sparse and may contain erroneous values. Therefore, this branch has hourly aggregation, missing data handling and outlier removal before obtaining the final output. In order to facilitate the flexibility of the pipeline, users can also disable the default outlier removal or data imputation and apply a custom removal process on the raw data.

The Intervention table contains features regarding procedures performed, such as mechanical ventilation and vasoactive medications (norepinephrine, phenylephrine, epinephrine etc.). The intervention features are treated as binary variables, with a series of 1s indicating the start time and end time of the procedure during the stay (Figure 2B), which follows the practice in MIMIC-Extract [16]. A complete list of variables is in SI Table S4-S6.

## 2.2. Variable query

By default, our pipeline extracts all demographic variables (SI Table S4) from `mimic_icu.icustays`, `mimic_core.admissions` and `mimic_core.patients` tables for

MIMIC-IV. For eICU, these demographic variables derive from `eicu_crd.patient`. We also extract 17 comorbidities for each ICU stay from `mimic_hosp.diagnoses_icd` and `eicu_crd.diagnosis`, respectively. One critical difference between the 2 databases is that MIMIC-IV uses both the International Classification of Diseases 9th Revision (ICD_9) and ICD_10 version codes while eICU only uses ICD_9 code. Even when querying using the same ICD_9 standard, MIMIC-IV stores the code a little differently. For instance, congestive heart failure could be represented by ICD codes 39891, 40201 etc. in MIMIC-IV, while in eICU, it's recorded as 398.91, 402.01 etc. We paid extra caution in designing the query code to miss as few comorbidity-related records as possible.

The Vital table variables contain time-varying measurements on blood gas, vital sign, urine chemistry etc. For MIMIC-IV, these variables were queried from `mimic_hosp.labevents`, `mimic_icu.chartevents` and `mimic_icu.outputevents`. For eICU, the raw results came from `eicu_crd.lab`, `eicu_crd.nursecharting`, `eicu_crd.intakeoutput`, `eicu_crd.microlab`, `eicu_crd.vitalperiodic`, `eicu_crd.respiratorycharting`. We also made use of existing derived tables from both repositories [22,23] to develop our SQL queries. There are in total 15 variables that were not found in the eICU database and we placed all NAN values in order to have the same dataframe shape for model cross validation. A list of these variables is in SI Table S7. In order to facilitate infectious disease research utilizing METRE, we extracted variables related to the antibiotics administration and microculture for both databases. Culture sites from MIMIC-IV have 57 unique values while the eICU database recorded 20 unique culture sites. These culture sites lack simple one-to-one correspondence, so we grouped the results into 14 categories based on the semantics. The details on creating concept-mapped microculture-related output variables are in SI Section 4.

Different from the Vital table where we focus on the numerical values for most variables, the Intervention table queries the start time and the end time of every intervention procedure. Besides ventilation, vasoactive agent, colloid bolus and crystalloid bolus, which are included in MIMIC-Extract [16], we further queried CRRT and 3 different types of transfusion as well as antibiotics administration. Another distinct difference between MIMIC-IV and eICU databases is that in MIMIC-IV, each measurement/treatment is associated with a unique `itemids`, while in eICU, it's usually directly represented by a unique string. The query strategy designed to obtain intervention-related records from both databases is in SI section 4.

## 2.3. Post processing

We performed post-processing on the Vital table and the Intervention table. For the Vital table, after getting the raw entries of each feature, we aggregated the feature by hour (Figure 2A) since most of the variables were not frequently recorded. For example, in MIMIC-IV, on average, heart rate was recorded every 1.07 hours and troponin-t was only recorded only every 131.46 hours. Users could choose their own time window ranging from 1h to 24h. After this aggregation, certain time points will have missing values. Before we implemented any imputation algorithm, we added a binary indicator column for each numerical variable (1 indicating the value is recorded and 0 indicating the value is imputed), since the recorded values could have higher credibility compared with the imputed values. Also noted by Ghassemi et al. [14], learning models without an appropriate model of missingness leads to brittle models when facing changes in measurement practices. Users have the option to discard the indicator column variables for a light-weight end-result. We then checked for outliers in the numerical variables. We made use of the list from in the source code repository of Harutyunyan et al. [24], which was based on clinical experts' knowledge of valid clinical measure ranges. For any value below the `outlier_low` or above the `outlier_high`, we emptied that cell and set the corresponding indicator cell to 0. Importantly, the same outlier removal criteria were applied on both MIMIC-IV and eICU to prevent introducing bias at this stage. Variable filling information before and after the outlier removal is in SI Table S1-S3. All the ranges used in the outlier removal are in SI Table S9. We also compared the mean and std in each variable between MIMIC-IV and eICU and demonstrated variables with the largest and smallest mean value difference (SI Figure S1, S2).

For the Intervention table, based on the start and end time of each variable, we added 1 or 0 indicating whether the treatment was performed in each hour (Figure 2B). There was no special missing value imputation in this table and 0 indicates that treatment was not performed in that hour. The default pipeline also has a variable reordering step making sure eICU dataframes have the same ordering as MIMIC-IV dataframes, which facilitates direct cross-validation between these 2 databases.

Before we implemented the default data imputation, we checked the ratio of null values for each ICU stay and for each variable. Users can set an optional missingness threshold above which means a stay or variable is not well-documented and will be removed.

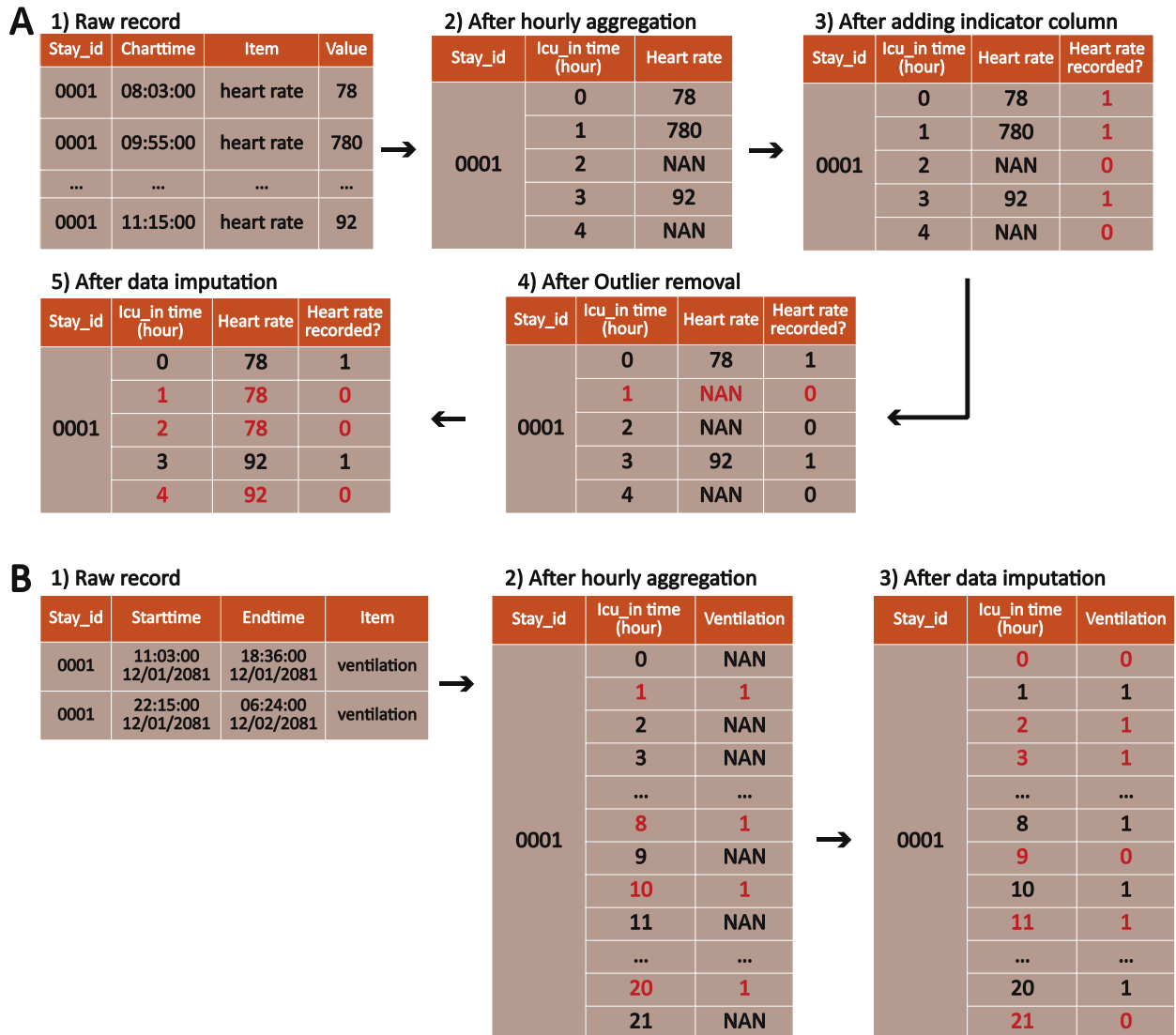

Figure 2. Post processing done on time-series variables with one-time entry and variables spanning hours. Notably, we added a binary indicator column for each numerical variable in A) to explicitly model the data missingness. A) the Vital table. The hypothetic patient was admitted to the ICU at 8:00 and 780 bpm is treated as an outlier for heat rate measurement. B) the Intervention table. The hypothetic patient was admitted to the ICU at 10:00 12/01/2081.

## 2.4. Baseline tasks

In order to demonstrate the utility of METRE, we extracted the eICU and MIMIC-IV data using the default choices and performed a number of clinically relevant prediction tasks using different model architectures.

**2.4.1. Tasks:** We incorporated 5 prediction tasks as defined by Tang et al. [18], which are:

1) In-hospital mortality prediction. In this task, 48h of extracted data is used for each stay. ICU stays with LOS < 48h are excluded. The prediction target is binary and the ground truth is from `hospital_expire_flag` in `mimic_core.admissions` for MIMIC-IV and for eICU, `eicu_crd.patients` record `unitdischargestatus` ('Alive' or 'Expired') explicitly.

    2-3) Acute respiratory failure (ARF), using either 4h (task 2) or 12h (task 3) of data. ARF is identified by the need for respiratory support with positive-pressure mechanical ventilation. For both MIMIC-IV and eICU, ventilation and positive end-expiratory pressure have been queried as variables so labels can be directly generated for each record.

    4-5) Shock, using either 4h (task 4) or 12h (task 5) of data. Shock is identified by receipt of vasopressor including norepinephrine, epinephrine, dopamine, vasopressin or phenylephrine. They have also been stored in the extracted Intervention table. For ARF and shock, the onset time is defined as the earliest time when the criteria is met.

Incorporating a gap between the end of the observation window (48h, 4h or 12h) and the onset of the positive target (time of death, ARF onset, shock onset) prevents data leakage and more closely resembles the real use case, where the care team takes possible measures after the model gives an alert. We incorporate gap hour into METRE, where positive cases are those whose onset time is observed during the ICU stay but out of observation window and the gap hour window. The negative cases are defined as no onset during the entire stay. Therefore, each task has a distinct study cohort. We used 6h as the gap hour but users can set their preferrable gap hours (including no gap). We also performed the same series of predictions without gap hour and the results are in SI Table S12-S14.

**2.4.2. Models**: We compared 4 different modeling approaches.

    1) Logistic regression (LR) models. For LR models, we used the sklearn linear model library and applied both L1 and L2 penalties [25]. Baysian optimization [26] was used for tuning the inverse of regularization strength C and the ElasticNet mixing parameter l1_ratio in order to maximize the average area under the receiver operating characteristic curve (AUC).

    2) Random forest (RF) models. RF models were built using sklearn ensemble library [27]. 6 hyperparameters including the number of trees in the forest, the maximum depth of the tree, the minimum number of samples required to split an internal node, the minimum number of samples required to be at a leaf node, the number of features to consider when looking for the best split, the number of samples to draw from the train set to train each base estimator were optimized using Bayesian optimization, with the same goal of maximizing AUC. For both LR and RF models, Bayesian optimization was implemented for 10 iterations with 5 steps of random exploration with expected improvement as the acquisition function. Besides, for both LR and RF

models, since it requires 1D input, we flattened the time-series data before feeding into the model.

      3) 1-dimensional convolutional neural networks (CNN) [28]. For CNN models, a random search on the convolutional kernel size, layer number, filter number, learning rate and batch size with a budget of 5 was implemented to maximize the AUC.

      4) Long short-term memory networks (LSTM) [29]. We also used the same number of random search budget for the number of features in the LSTM hidden state, the number of recurrent layers, the feature dropout rate, learning rate and batch size.

      The train-test split is 80:20 and the train set is used in a 10-fold cross validation. The test set AUC and area under the precision-recall curve (AUPRC) performance was reported. Empirical 95% confidence intervals were also reported using 1000 bootstrapped samples of the test set and under 1000 bootstraps. Code to run these models is available online and more detailed parameter choices on the modeling are in the SI Section 9.

## 3. RESULTS

We first compared METRE with other MIMIC extraction pipelines (Table 1). METRE is the only pipeline that extracts the newest MIMIC database MIMIC-IV and eICU in a consistent way, includes a diverse set of EHR variables, and at the same time allows for both generic and condition-specific cohort selection. Notably, FIDDLE makes use of the underlying data distribution and performs post-processing similar to one-hot encoding, which results in the creation of large numbers of synthetic time-series variables. We also acknowledge that using MIMIC-III data has the benefit of abundant references and consequent in-depth understanding of the database.

*Table 1. Comparison of METRE with prior works*

|  | METRE | MIMIC-Extract [16] | FIDDLE [18] | Gupta et al. [30] | Cop-e-cat [17] |
|---|---|---|---|---|---|
| Time-series variable number | 108 | 104 | A few hundred to a few thousand depending on the task | NA | 43 |
| If used MIMIC, MIMIC-III or IV? | IV | III | III | IV | IV |
| Included medication and | Y | Y | Y | Y | Y |
| procedure variables? | | | | | |
| Generic cohort | Y | Y | Y | Y | Y |
| Condition-specific cohort? | Y | N | Y | Y | N |
| Extracted eICU for cross validation? | Y | N | N, extracted eICU, but not consistent with MIMIC-IV | N | N |
| Flexible time Window? | Y | N | Y | Y | Y |

Table 2 is the demographic and ICU stay summary of the extracted MIMIC-IV and eICU cohorts using the default settings (generic cohort, age>18, data missingness threshold 0.9, LOS> 24h and LOS < 240h). Users can obtain their custom cohorts by defining the keyword, age constraint, LOS length constraint and by choosing whether to remove poorly populated stays or not. The age, gender, ethnicity distributions between these 2 cohorts are very similar.

*Table 2. Demographic and ICU stay summary of the default MIMIC-IV and eICU cohort in METRE*

| | | MIMIC-IV (N = 38766) | eICU (N = 126448) |
|---|---|---|---|
| Gender | Female | 16825 (43.4%) | 58028 (45.9%) |
| | Male | 21941 (56.6%) | 68339 (54.0%) |
| | Other/Unknown | 0 | 81 (0.1%) |
| Age | < 30 | 1753 (4.5%) | 6035 (4.8%) |
| | 31- 50 | 5264 (13.6%) | 18782 (14.9%) |
| | 50 – 70 | 15748 (40.6%) | 52590 (41.6%) |
| | > 70 | 16001 (41.3%) | 49041 (38.8%) |
| | | mean: 65.1, std: 16.8 | mean: 63.9, std:16.6 |
| Ethnicity | American Indian/Alaska Native | 63 (0.2%) | 844 (0.7%) |
| | Asian | 1137 (2.9%) | 2087 (1.7%) |
| | Hispanic | 1319 (3.4%) | 4651 (3.7%) |
| | Black/African American | 3384 (8.7%) | 14013 (11.1%) |
| | Other/Unknown | 6592 (17.0%) | 7142 (5.6%) |
| | White | 26271 (67.8%) | 97711 (77.3%) |
| Admission Year (Discharge Year in eICU) | 2008 – 2010 | 12028 (31.0%) | 2014: 60002 (47.4%) |
| | 2011 – 2013 | 9393 (24.2%) | 2015: 66446 (52.6%) |
| | 2014 - 2016 | 9284 (23.9%) | |
| | 2017 - 2019 | 8061 (20.8%) | |

|  |  | **MIMIC-IV (N = 38766)** | **eICU (N = 126448)** |
|---|---|---|---|
| Hospital Mortality | No<br>Yes<br>Unknown | 35214 (90.8%)<br>3552 (9.2%)<br>0 | 114912 (90.1%)<br>10438 (8.3%)<br>1098 (0.9%) |
| ICU Mortality | No<br>Yes<br>Unknown | 36518 (94.2%)<br>2248 (5.8%)<br>0 | 120518 (95.3%)<br>5923 (4.7%)<br>7 |
| ICU Stay Duration | 1 - 3 Days<br>3 - 7 Days<br>7 - 10 Days | 26010 (67.1%)<br>10515 (27.1%)<br>2241 (5.8%)<br>mean: 2.9, std: 1.0 | 85750 (67.8%)<br>34007 (26.9%)<br>6691 (5.3%)<br>mean: 2.8, std: 1.9 |

Using the default cohorts shown in Table 2, we first demonstrate the utility of METRE on the five tasks mentioned previously (and demonstrated in FIDDLE [18]) using 4 different training architectures (LR, RF, 1D CNN and LSTM). The complete results are in Table 3. We used AUC and AUPRC of the test set as evaluation metrics. For MIMIC data, models trained with data extracted using METRE achieved 5/10 best performance in the tasks performed, demonstrating the utility of the pipeline. One benefit of METRE is the built-in ability to extract from the eICU database in parallel; we used the same criteria to obtain ARF and shock labels for eICU cohorts and then compared different model performance with those listed in FIDDLE. Models trained using data from the METRE-eICU also have comparable performance with those developed in FIDDLE, as expected (SI Table S10).

Next, we evaluated the AUC and AUPRC from the unseen dataset for models trained on the other dataset. For the hospital mortality prediction model trained on MIMIC-IV, when we directly evaluated its performance on the whole eICU data (not only eICU test set, so the cohort size here is 60468) without any fine tuning, the best AUC is 0.824, which was achieved by the LSTM model. This only presents a 0.044 drop from the original MIMIC-IV LSTM in-hospital mortality model AUC of 0.868. Similarly, when we reversed the role of these 2 databases, the best AUC is 0.833, which is only 0.013 lower compared to the original eICU test set performance 0.846. Figure 3 compares the MIMIC-IV/eICU cross validation AUC on all 8 hospital mortality predictions, which clearly demonstrates the utility that arises from facile multidatabase extraction. **Error! Reference source not found.** has a complete comparison between the external dataset validation performance across models and across 5 different prediction targets. Among all 5 tasks, the ARF prediction task presented an AUC drop as high as 0.322 when the model was trained using eICU and evaluated using MIMIC-IV. We hypothesize that the generation of the ARF label may be dependent on local clinical care practice (e.g., whether the patient was on ventilation or not) and the completeness of the entries into the database. We therefore compared the ratio of people receiving all the 16 intervention procedures in SI Table S15 and observed the differences between MIMIC-IV and eICU. Model generalization performance has been known to be a challenge in model deployment, especially in clinical contexts, where the incoming data is often generated under different care practices. Therefore, we believe the ability of our pipeline to perform model testing using data collected from different institutions prior to deployment is of critical importance.

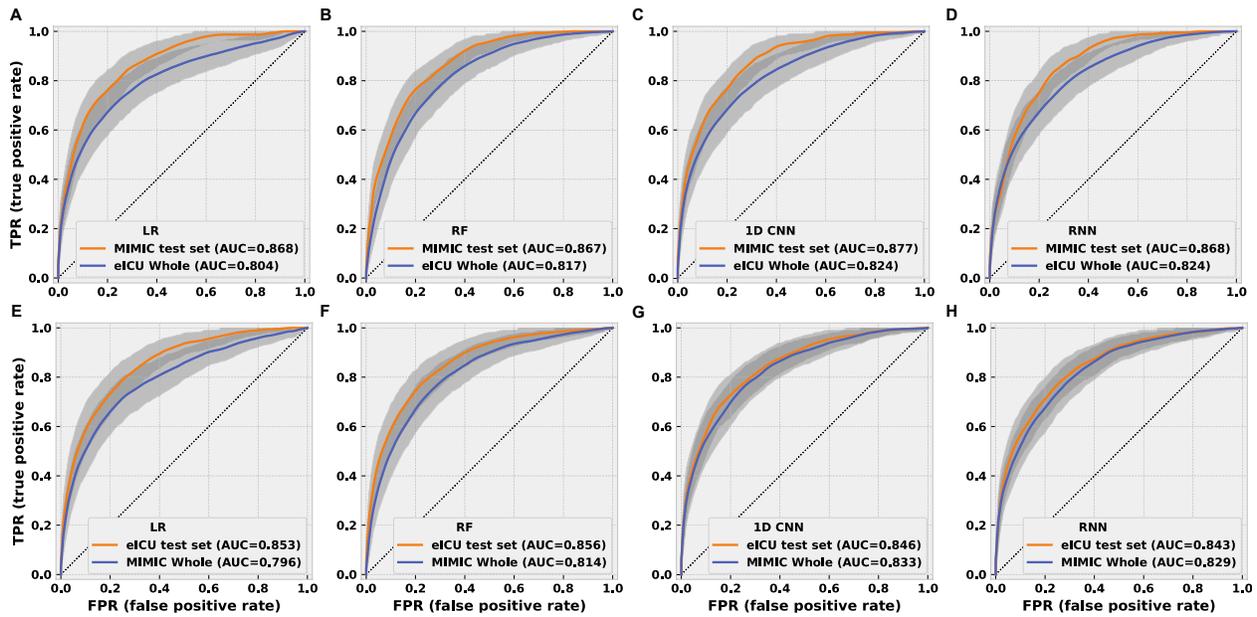

Figure 3. Cross validation performances on in-hospital mortality prediction. A) LR, B) RF, C) CNN, D) LSTM models were trained using tables derived from MIMIC-IV and tested using both MIMIC-IV test set and the whole eICU set. E) LR, F) RF, G) CNN, H) LSTM models were trained using tables derived from eICU and tested using both eICU test set and the whole MIMIC-IV set. AUC curves with 95% confidence interval are shown in each panel.

Table 3. AUC and AUPRC comparison of trained models on the 5 different tasks. Data is from MIMIC.

| | | Hospital Mortality 48h | | ARF 4h | | ARF 12h | | Shock 4h | | Shock 12h | |
|---|---|---|---|---|---|---|---|---|---|---|---|
| | | AUROC | AUPRC | AUROC | AUPRC | AUROC | AUPRC | AUROC | AUPRC | AUROC | AUPRC |
| MIMIC Extract (MIMIC-III) | LR | 0.859 (0.830-0.887) | 0.445 (0.358-0.540) | 0.777 (0.752-0.803) | 0.604 (0.561-0.648) | 0.723 (0.683-0.759) | 0.250 (0.200-0.313) | 0.796 (0.771-0.821) | 0.505 (0.454-0.557) | 0.748 (0.712-0.784) | 0.242 (0.193-0.310) |
| | RF | 0.852 (0.821-0.882) | 0.448 (0.359-0.537) | 0.821 (0.799-0.843) | 0.660 (0.617-0.698) | 0.747 (0.709-0.782) | 0.289 (0.235-0.356) | 0.824 (0.801-0.845) | 0.541 (0.488-0.588) | 0.778 (0.742-0.812) | 0.307 (0.248-0.369) |
| | CNN | 0.851 (0.820-0.879) | 0.439 (0.353-0.529) | 0.788 (0.763-0.814) | 0.633 (0.591-0.672) | 0.722 (0.684-0.758) | 0.258 (0.207-0.320) | 0.798 (0.773-0.824) | 0.520 (0.471-0.572) | 0.741 (0.704-0.778) | 0.247 (0.198-0.317) |
| | LSTM | 0.837 (0.803-0.867) (n = 1264) | 0.441 (0.358-0.523) | 0.796 (0.770-0.822) (n = 2358) | 0.634 (0.590-0.675) | 0.700 (0.661-0.736) (n = 2093) | 0.229 (0.184-0.286) | 0.801 (0.778-0.825) (n = 2867) | 0.513 (0.463-0.562) | 0.753 (0.717-0.791) (n = 2612) | 0.248 (0.199-0.313) |
| FIDDLE (MIMIC-III) | LR | 0.856 (0.821-0.888) | 0.444 (0.357-0.545) | 0.817 (0.792-0.839) | 0.657 (0.614-0.696) | 0.757 (0.720-0.789) | 0.291 (0.236-0.354) | 0.825 (0.803-0.846) | 0.548 (0.501-0.595) | 0.792 (0.758-0.824) | 0.274 (0.227-0.338) |
| | RF | 0.814 (0.780-0.847) | 0.357 (0.279-0.448) | 0.817 (0.795-0.839) | 0.652 (0.608-0.690) | 0.760 (0.726-0.793) | 0.317 (0.255-0.382) | 0.809 (0.786-0.833) | 0.516 (0.467-0.566) | 0.773 (0.740-0.806) | 0.288 (0.231-0.355) |
| | CNN | **0.886** (0.854-0.916) | 0.531 (0.434-0.629) | **0.827** (0.803-0.848) | **0.666** (0.626-0.705) | 0.768 (0.733-0.800) | 0.294 (0.238-0.361) | 0.831 (0.811-0.851) | 0.541 (0.493-0.589) | 0.791 (0.758-0.823) | 0.295 (0.239-0.361) |
| | LSTM | 0.868 (0.835-0.897) (n = 1264) | 0.510 (0.411-0.597) | 0.827 (0.801-0.846) (n = 2358) | 0.664 (0.623-0.703) | **0.771** (0.737-0.802) (n = 2093) | 0.326 (0.267-0.397) | 0.824 (0.803–0.845) (n = 2867) | **0.541** (0.497-0.587) | 0.792 (0.759-0.823) (n = 2612) | 0.314 (0.251-0.386) |
| METRE (MIMIC-IV) | LR | 0.868 (0.835-0.901) | 0.535 (0.444-0.625) | 0.730 (0.696-0.765) | 0.539 (0.477-0.601) | 0.723 (0.680-0.767) | 0.411 (0.339-0.484) | 0.763 (0.700-0.825) | 0.227 (0.135-0.319) | 0.769 (0.696-0.841) | 0.186 (0.081-0.291) |
| | RF | 0.867 (0.836-0.898) | 0.503 (0.408-0.597) | 0.779 (0.746-0.811) | 0.656 (0.606-0.707) | 0.756 (0.713-0.799) | **0.523** (0.452-0.595) | 0.880 (0.835-0.926) | 0.505 (0.385-0.625) | **0.888** (0.832-0.944) | **0.471** (0.090-0.402) |
| | CNN | 0.876 (0.846-0.906) | **0.550** (0.458-0.642) | 0.756 (0.720-0.792) | 0.624 (0.568-0.680) | 0.748 (0.707-0.790) | 0.497 (0.424-0.569) | 0.798 (0.740-0.856) | 0.302 (0.190-0.414) | 0.822 (0.757-0.887) | 0.279 (0.149-0.410) |
| | LSTM | 0.869 (0.837-0.900) (n=3786) | 0.505 (0.411-0.599) | 0.768 (0.734-0.802) (n=1973) | 0.639 (0.587-0.690) | 0.741 (0.700-0.782) (n=1740) | 0.477 (0.405-0.550) | **0.838** (0.786-0.890) (n=5451) | 0.405 (0.285-0.525) | 0.797 (0.723-0.871) (n=5321) | 0.256 (0.128-0.384) |

*Table 4. Cross validation performance (DIFF: AUC/AUPRC difference compared to the original test set)*

|  |  | Hospital Mortality 48h | | ARF 4h | | ARF 12h | | Shock 4h | | Shock 12h | |
|---|---|---|---|---|---|---|---|---|---|---|---|
|  |  | AUROC | AUPR | AUROC | AUPR | AUROC | AUPR | AUROC | AUPR | AUROC | AUPR |
| eICU validation (models trained on MIMIC-IV) | LR | 0.802 (0.754-0.850) | 0.460 (0.363-0.557) | 0.714 (0.627-0.802) | 0.140 (0.063-0.217) | 0.726 (0.621-0.831) | 0.126 (0.044-0.208) | 0.652 (0.565-0.738) | 0.111 (0.047-0.175) | 0.657 (0.549-0.765) | 0.090 (0.020-0.161) |
|  | DIFF | -0.066 | -0.075 | **-0.016** | -0.399 | **0.003** | -0.285 | -0.111 | **-0.116** | -0.112 | **-0.096** |
|  | RF | 0.817 (0.778-0.856) | 0.384 (0.293-0.475) | 0.804 (0.747-0.862) | 0.154 (0.079-0.230) | 0.791 (0.727-0.854) | 0.109 (0.041-0.177) | 0.782 (0.708-0.856) | 0.186 (0.092-0.280) | 0.776 (0.686-0.865) | 0.140 (0.050-0.231) |
|  | DIFF | -0.050 | -0.119 | 0.025 | -0.502 | 0.035 | -0.414 | -0.098 | -0.319 | -0.112 | -0.331 |
|  | CNN | 0.822 (0.779-0.865) | 0.463 (0.365-0.560) | 0.821 (0.748-0.893) | 0.236 (0.130-0.342) | 0.802 (0.713-0.891) | 0.185 (0.075-0.295) | 0.706 (0.622-0.789) | 0.131 (0.055-0.207) | 0.683 (0.573-0.793) | 0.093 (0.025-0.160) |
|  | DIFF | -0.054 | -0.087 | 0.065 | -0.388 | 0.054 | -0.312 | **-0.092** | -0.171 | -0.139 | -0.186 |
|  | LSTM | 0.824 (0.783-0.864) | 0.453 (0.360-0.546) | 0.894 (0.837-0.951) | 0.383 (0.242-0.525) | 0.828 (0.739-0.918) | 0.260 (0.113-0.408) | 0.722 (0.641-0.804) | 0.153 (0.065-0.242) | 0.706 (0.605-0.806) | 0.107 (0.028-0.186) |
|  | DIFF | **-0.045** | **-0.052** | 0.126 | **-0.256** | 0.087 | **-0.217** | -0.116 | -0.252 | **-0.091** | -0.149 |
|  |  | (n = 60468) | | (n = 97067) | | (n = 95927) | | (n = 110406) | | (n = 108953) | |
| MIMIC-IV validation (models trained on eICU) | LR | 0.796 (0.752-0.841) | 0.448 (0.355-0.540) | 0.664 (0.624-0.704) | 0.470 (0.408-0.533) | 0.666 (0.620-0.712) | 0.360 (0.288-0.433) | 0.688 (0.621-0.754) | 0.184 (0.112-0.256) | 0.700 (0.621-0.779) | 0.130 (0.068-0.192) |
|  | DIFF | -0.060 | -0.056 | **-0.207** | 0.190 | **-0.201** | 0.129 | -0.052 | **0.034** | -0.044 | **0.004** |
|  | RF | 0.813 (0.772-0.854) | 0.428 (0.339-0.517) | 0.667 (0.627-0.707) | 0.525 (0.469-0.581) | 0.678 (0.632-0.724) | 0.428 (0.361-0.496) | 0.841 (0.796-0.887) | 0.344 (0.236-0.451) | 0.846 (0.791-0.900) | 0.304 (0.178-0.431) |
|  | DIFF | -0.044 | -0.086 | -0.322 | -0.272 | -0.309 | -0.287 | **0.019** | 0.070 | **0.028** | 0.058 |
|  | CNN | 0.833 (0.794-0.872) | 0.510 (0.418-0.601) | 0.691 (0.650-0.732) | 0.549 (0.492-0.606) | 0.665 (0.617-0.713) | 0.435 (0.366-0.504) | 0.738 (0.675-0.801) | 0.229 (0.143-0.315) | 0.755 (0.684-0.825) | 0.195 (0.098-0.292) |
|  | DIFF | **-0.015** | 0.005 | -0.269 | **0.005** | -0.292 | -0.033 | -0.052 | 0.035 | -0.036 | 0.030 |
|  | LSTM | 0.830 (0.790-0.869) | 0.498 (0.405-0.591) | 0.704 (0.664-0.743) | 0.573 (0.517-0.629) | 0.682 (0.637-0.726) | 0.431 (0.363-0.498) | 0.749 (0.690-0.808) | 0.245 (0.155-0.336) | 0.751 (0.678-0.825) | 0.204 (0.102-0.306) |
|  | DIFF | -0.015 | 0.005 | -0.277 | -0.150 | -0.278 | **-0.126** | -0.040 | 0.036 | -0.040 | 0.052 |
|  |  | (n = 18815) | | (n = 9713) | | (n = 8594) | | (n = 27232) | | (n = 26546) | |

## 4. DISCUSSION

METRE bridges the gap between MIMIC-IV and eICU by creating harmonized outputs that allow for facile cross-validation across the two datasets, which greatly benefits 1) Users who want to evaluate their model on a dataset that hasn't been seen during the development process. 2) Users who want to model larger cohorts; the combined cohort size of MIMIC-IV and eICU in METRE is 165254.

METRE was developed to allow for substantial user flexibility. 1) Users can specify their own cohort choices on age, LOS, record missingness threshold at the beginning of the pipeline. 2) Users can also choose to stop at a few exit points defined at the pipeline. The exit points allow users to get outputs before implementing the default imputation method, or before normalizing the output, or before performing the train-validation-test split. In this way, users can merge their preferred design choices into the pipeline. 3) We only provided a limited set of arguments for condition-specific cohorts. To accommodate this, users can first locate the `stayid` (MIMIC-IV) or the `patientunitstayid` (eICU) for their specific cohort and use our pipeline under `customid` argument to do the rest of the variable query and cleaning work.

There are some interesting METRE variables that we didn't explore in this work. For instance, in our MIMIC-IV data, we extracted patient discharge year as a static feature. It has been studied by Nestor et al [21] that date-agnostic models could overestimate prediction quality and affect future deployment potential. We hope that METRE could spark research work exploring this feature and developing models that generalizes better over changing health care practices.

Although METRE's key value lies in the flexibility in extracting the data, to demonstrate the effectiveness of the pipeline, we performed 5 clinically relevant prediction tasks with hospital mortality, ARF, shock as the target and developed LR, RF, CNN and LSTM models for each task. With AUC and AUPRC as the metrics, all the models have comparable performance with models developed using MIMIC-Extract and FIDDLE, as expected. However, our pipeline has the additional advantage of enabling facile cross validation between MIMIC-IV and eICU datasets. It is also worth noting that we are not attempting to develop a one-size-fits-all solution that generalizes across all databases; this is an enduring challenge in the EHR field. METRE aims at expediting the data preprocessing stage for researchers who are interested in using both MIMIC-IV and eICU data. We welcome community contributions to METRE to keep developing additional functionality.

## 5. CONCLUSION

We developed an open-source cohort selection and pre-processing pipeline METRE to extract multi-variate EHR data. We focused on 2 widely-used EHR databases: MIMIC-IV and eICU. METRE produces a wide variety of variables including time-series variables such as labs, vital signs, treatments, and interventions as well as static variables including demographics and comorbidities. Our open-source pipeline transforms MIMIC-IV and eICU into structured data frames and allows researchers to perform model testing using data collected from different institutions, which is of critical importance for model deployment under clinical context.

## 6. DATA AVAILABILITY

The code used to extract the data and perform training is available here:
https://github.mit.edu/voldman-lab/METRE_MIMIC_eICU

## 7. ACKNOWLEDGEMENTS

We thank Dr. Luca Daniel (Massachusetts Institute of Technology), Dr. Tsui-Wei Weng (University of California San Diego), Ching-Yun Ko (Massachusetts Institute of Technology) for valuable discussions.

## 8. FUNDING

The project is funded by MIT Jameel Clinic for Machine Learning in Health and W.L. received MIT EWSC Fellowship while conducting the research.